# Measurement of real contact area for rough metal surfaces and the distinction of contribution from elasticity and plasticity

Lei-Tao Li, Xuan-Ming Liang, Yu-Zhe Xing, Duo Yan, Gang-Feng Wang[*]

*Department of Engineering Mechanics, SVL, Xi'an Jiaotong University, Xi'an 710049, China*

[*] Corresponding author. wanggf@mail.xjtu.edu.cn

**Abstract:** The measurement of real contact area between rough surfaces is one of most challenge problems in contact mechanics and is of importance to understand some physical mechanisms in tribology. Based on the frustrated total internal reflection, a new apparatus is designed to measure the real contact area. For brass samples with various surface topographies, the relation between normal load and real contact area is measured. Our experiments verify the linear relationship between load and real contact area during loading process. More importantly, the unloading process is firstly considered to distinguish the contribution of elasticity and plasticity in contact of rough surfaces. It is found that both elasticity and plasticity are involved throughout the continuous loading process, different from some present understanding and assumptions that they play at different loading stages. A quantitative parameter is proposed to indicate the contribution of plasticity. The present work not only provides an experimental method to measure real contact area but figure out how elastic and plastic deformation works in contact of rough surfaces.

# 1. Introduction

For contact of rough surfaces, the real contact area (RCA) is only a small fraction of the apparent area, but plays an important role in friction, wear and sealing. For examples, it is usually supposed that the friction force is proportional to the real contact area [1], and the sealing effect is determined by the real contact surface [2]. Therefore, much effort has been devoted to revealing the relationship between load and RCA.

The contact of rough surfaces happens in fact at a collection of discrete contact spots, and various multi-asperity models have been developed. In the pioneering work of Greenwood and Williamson [3], the height of asperities follows the Gaussian distribution and their curvature radius is identical. Then using the spherical Hertzian solution, an approximately linear relation between load and RCA is predicted. Archard [4] described rough surfaces by protuberances covered with even smaller protuberances and so on, and gave a general power-law relationship between load and RCA depending upon the form of surfaces. A general self-affine fractal contact model is further developed by Majumdar and Bushan [5]. Using the magnification referring to some chosen length scale, Persson [6] formulated a contact model holding up to nearly complete contact. Interaction among asperities is also an important factor and should be taken into account [7-10]. Since contact sizes cross various length scales, some microscopic mechanisms, such as strain gradient at micrometers [11] and surface effect at nanometers [12-13], are incorporated in contact models.

In the GW model, the deformation of asperities is entirely elastic. While on the other hand, Bowden and Tabor assumed that each contact asperity deforms permanently by plastic flow. Later, Greenwood and Wu [14] pointed out that contact deformation may be plastic at light load and become elastic at heavy load. The contact state of asperity, whether it deforms elastically or plastically, is of considerable importance to safe sliding [15]. However, up to now, scare experimental work has been conducted to distinguish the deformation state in elasticity or plasticity and their contribution in contact of rough surfaces.

Experiments have been conducted to measure the real contact area using electrical, optical [16], and ultrasonic techniques [17]. Among these approaches, optical method allows for direct observation of real contact area and has gained many applications in recent years. Using a monolayer of fluorescent molecules, Weber et al [18-19] measured the real contact area in-situ and verified the linear relationship between friction force and real contact area. Through an in-situ tribometer, Lechthaler et al [20] measured the evolution of real contact area under cyclic sliding, in which the real contact area was determined numerically from height information. Using the total internal reflection method, Song et al [21] measured the real contact area between two transparent PMMA blocks, in which both optical refection and transmission were required. Based on the frustrated total internal reflection (FTIR), Bennett et al [22] measured the real contact area between a rough opaque PMMA and a smooth transparent PDMS with increasing pressure. The experimental measurements showed a close agreement with the simulation results, displaying the efficiency of FTIR technique in contact measurements.

In this work, the FTIR method is employed to measure the real contact area of rough metal surfaces. To distinguish the contribution of elasticity and plasticity, both loading and unloading are considered. In the continuous loading process, a linear relation between load and RCA is observed, in which both elastic and plastic deformation are involved. While in an unloading process with only elastic deformation, a nonlinear relationship is firstly noticed. For metals with various topographies, such characteristics are repeatable.

## 2. Experiment

### 2.1 Sample preparation

In order to investigate the influence of surface topographies, various cubic samples are prepared as shown in Fig. 1. Sample T1, T2, and T3 are made of brass (4 mm × 4 mm × 5 mm), while T4 is made of copper (4 mm × 5 mm × 5 mm). The surface of T1 is produced by making micro-pits randomly with a micro-hand grinding. The surface of T2 is polished in one direction by using a surface grinder, which is common in machining. Sample T3 is rolled to form a random rough surface. The surface of T4 is firstly cut into a zigzag shape by linear cutting machine and then flattened by a grinder to form a corrugated rough surface. The material properties are presented in Table 1. Five to ten samples of each type were prepared for repeated tests.

### 2.2 Experimental apparatus

Based on the FTIR technique, we design a self-made apparatus to measure the real contact area of rough surface, and its schematics is presented in Fig. 2. The experimental apparatus includes three parts: the FTIR contact device, a loading system and a CCD camera (SHUNHUALI).

The load system is provided by a synchronous, ball screw and force sensor (STB671, SIMTOUCH) mechanism, which provides a pressure ranging from 0 N to 1 kN at a constant loading rate in the vertical direction. In our experiments, the loading/unloading rate is set at 0.5 mm/min.

The key device in the FTIR contact plate is a quartz glass disc with a diameter of 200 mm and a thickness of 15 mm. The glass is much harder than the test cubes, therefore its upper surface will keep flat and be easy to focus on. The glass disc is circumferentially wrapped with a monochromatic purple LED strip, and they are fixed together through an aluminum clip. The LED strip and the edge of the glass disc are covered with a blackout fabric to ensure all light inside the glass incident at an angle larger than the total reflection critical angle. All experiments are conducted in a dark environment to eliminate the interference of stray light.

To measure the real contact area, a CCD camera is set up beneath the glass disc and its focus is adjusted on the contact surface between the glass disk and the test tube. At the region without real contact, total reflection will happen and light will be reflected back completely into the glass. While at the area of intimate contact, transmission will occur and the reflection will be less than total. Such difference can be captured by CCD below. In the process of taking pictures, we use the manual mode of the loading device, and keep the aperture, shutter speed and focal length of CCD unchanged. The images of contact surface are continuously captured by the CCD camera at a rate of 25 fps. After calibrated by a 10 × magnification lens, the pixel size is about $10 \times 10$ μm$^2$.

By using improved Otsu technique [21], images are transformed into binary images as shown in Figs. 3 and 4. Taking T2 and T3 as examples, it can be observed that the images are filled with light and dark spots. The light part is formed by light transmission on the real contact spots and the dark areas are attributed to the total internal reflection on the non-contact spots. To obtain the RCA accurately, a program was coded to count the number of contact points automatically. Then, the relationship between the RCA and the compression force can be obtained.

A cyclic load test was performed for each sample. First the sample was fixed to the indenter, and then the motor-driven indenter compressed the sample. Once the loading force reaches an assigned value, the motor reverses to unload until the compression force drops to zero. A loading and an unloading process form a complete loading cycle. After that, the indenter compresses the sample again to a higher load, and a new cycle begins.

## 3. Results and discussions
### 3.1 The relation between load and real contact area

For the sample T1, the development of RCA with respect to load is played in Fig. 5. If the load is continuously increased without unloading, the RCA is approximately proportional to the load, as shown by the dotted line. This result has been observed in previous experiments [4], and can be explained by some theoretical models [23]. But what is the contribution of elasticity and plasticity in the loading process is unknown. Therefore, the unloading is performed at several loads.

As an example, attention to the first unloading since $F = 55$ N in Fig. 5. As the load declines continuously till vanishing, the RCA decreases monotonously too but nonlinearly. In the unloading process, it is known that the deformation is purely elastic. Since the loading curve is distinct from the unloading one, it is concluded that both elastic and plastic deformations are involved in the monotonous loading process.

When a re-loading is conducted subsequently, the RCA will increase again along the former unloading curve till $F = 55$ N, since only elastic deformation happens in this range. However, when the load is beyond 55 N, the RCA increases linearly again with respect to load. Moreover, the slope of loading curve keeps approximately the same as the one before unloading happens. Such characteristics repeat in each loading and unloading cycle. The same experiments have been conducted on samples T3 and T4 with different topography, and similar results have been observed, as shown in Fig. 6 and Fig. 7.

## 3.2 The distinction of contribution from elasticity and plasticity

We summarize the above experimental results in a more general conclusion, as shown in Fig. 8. In the continuous loading process without any unloading interrupt, both elasticity and plasticity are involved in generating the RCA, which is linearly proportional to the load. While when an unloading is performed, which is related only to elasticity, the dependence of RCA on load is nonlinear. Such features can be employed to distinct the contribution of elasticity and plasticity in the contact of rough surfaces.

At a specific load, the straight loading curve and the bended unloading curve form a closed loop, which is owing to the plastic deformation in the loading process and thus can be used to indicate the contribution of plasticity. We use the area $S_p$ enclosed by the load-unloading curve to quantitatively represent the contribution of plastic deformation. While the contribution including both elasticity and plasticity is represented by the area $S_{ep}$ below the straight loading curve. Then the ratio $S_p$ to $S_{ep}$, named as plastic index, can indicate the relative contribution of plasticity to some extent.

For the results in Fig. 5, Fig. 6, Fig. 7 and the experimental result on sample T2, we calculate the plastic index as displayed in Fig. 9, in which $E^*=E/(1-v^2)$ and $A$ are the composite elastic modulus and the nominal contact area, respectively. It is seen that, for the considered metallic rough surfaces, both elasticity and plasticity play important role throughout the whole loading process. Surface morphologies may affect the relative contribution at light load. However, at large load, the relative contribution of plasticity converges almost to a constant, i.e. 0.5 for the considered metals.

4. Conclusions

Based on the FTIR technique, an apparatus has been designed to measure the real contact area of rough surfaces. The development of RCA with increasing load is measured, which verifies the linear relation between them. To distinguish the contribution from elastic and plastic deformations, unloading is conducted at several loads, which reveals the nonlinear dependence of RCA upon load at these stages. It is found that neither elasticity nor plasticity is dominant at any stage in a continuous loading process, but both of them play an important role. A plastic index is proposed to indicate the contribution of plasticity in the RCA. For the considered metallic surfaces, the contribution of plastic deformation keeps almost constant at heavy load independent of surface topographies. These results improve our understanding about the influence of elasticity and plasticity in rough surface contact.

**Acknowledgements**

Supports from the National Natural Science Foundation of China (Grant No. 11525209) are acknowledged.

Table 1

Material properties.

| Material | Brass | Copper |
|---|---|---|
| $E$: Young's Modulus/Gpa | 80 | 118 |
| $v$: Poisson's ratio | 0.37 | 0.34 |
| $H$: Hardness/HB | 59 | 37 |
| $\sigma_{0.2}$: Yield Strength/MPa | 33.3 | 33.3 |
| $\sigma_b$: Tensile Strength/MPa | 209 | 209 |

**Figure Captions**

Fig. 1. Test cubes T1, T2, T3, T4 and their surfaces

Fig. 2. Schematic diagram of experimental system to measure real contact area

Fig. 3. Real contact areas of sample T2 at different loading stages

Fig. 4. Real contact areas of sample T3 at different loading stages

Fig. 5. Evolution of real contact area for T1 unloaded at 55N, 165N, 252N, 370N and 460N,

Fig. 6 Evolution of real contact area for T3 unloaded at 100N, 125N, 148N, 175N and 200N,

Fig. 7 Evolution of real contact area for T4 unloaded at 50N, 150N, 250N, 350N and 450N

Fig. 8. Schematic distinction of elastic and plastic contribution in RCA/F curve

Fig. 9. Development of plastic contribution with respect to load

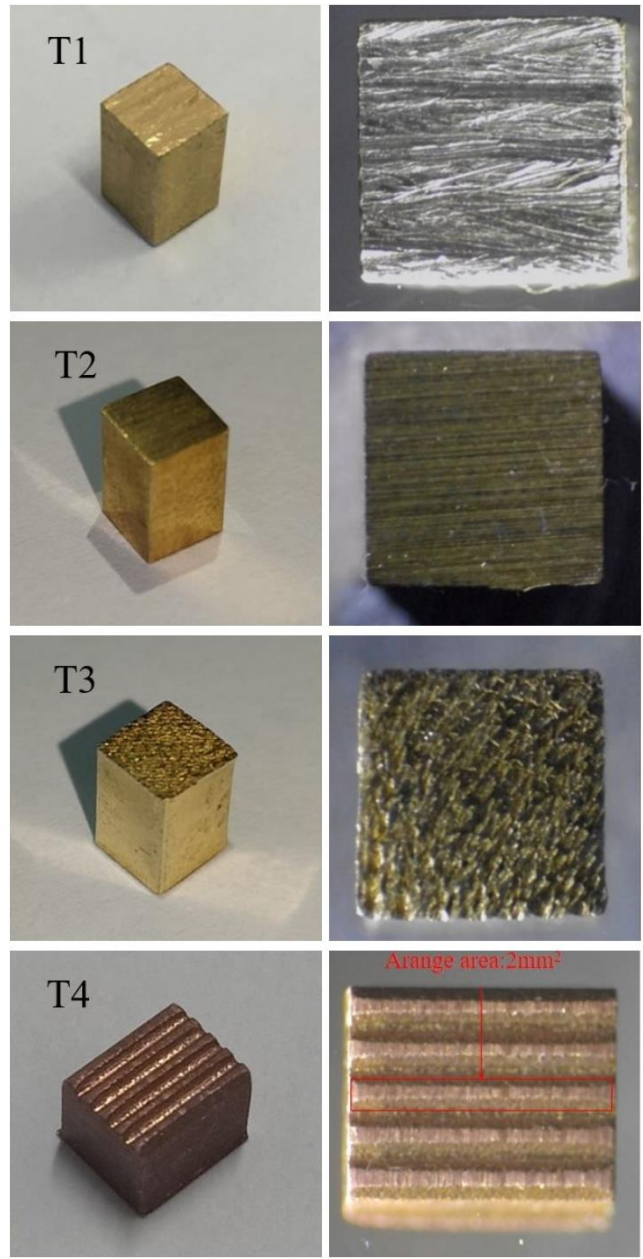

Fig. 1.

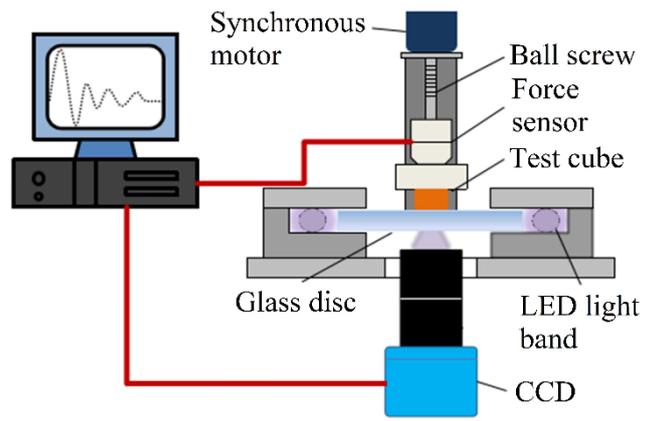

Fig. 2.

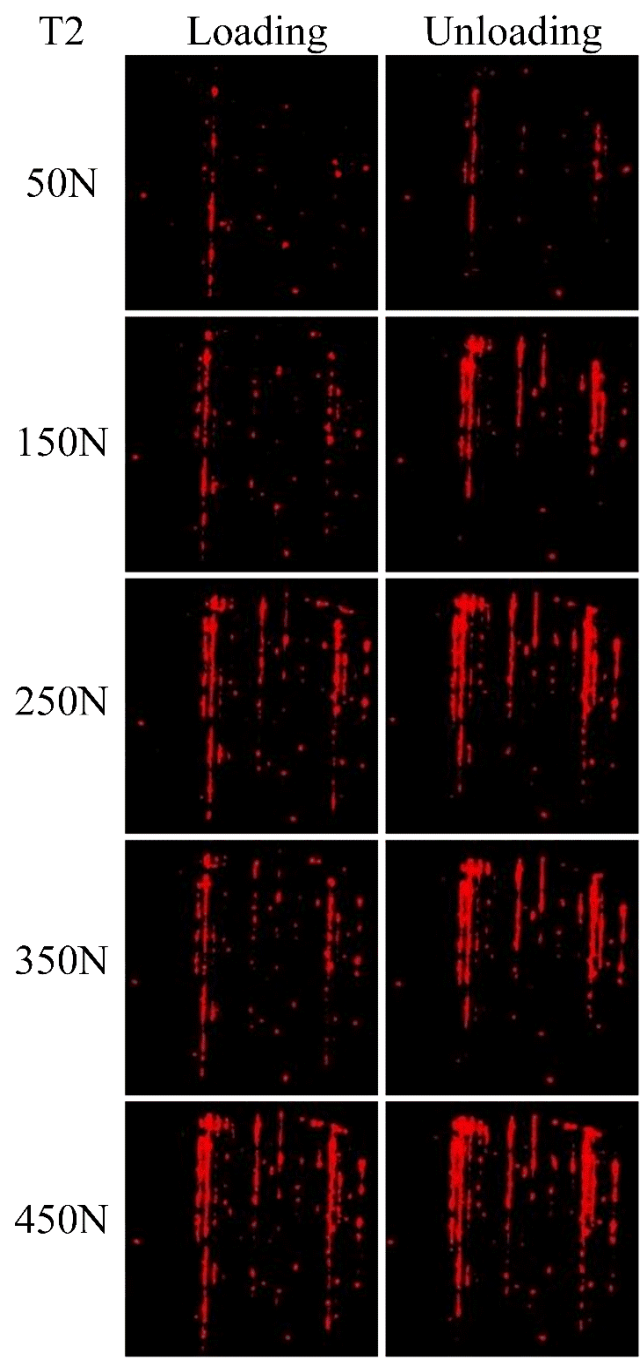

Fig. 3.

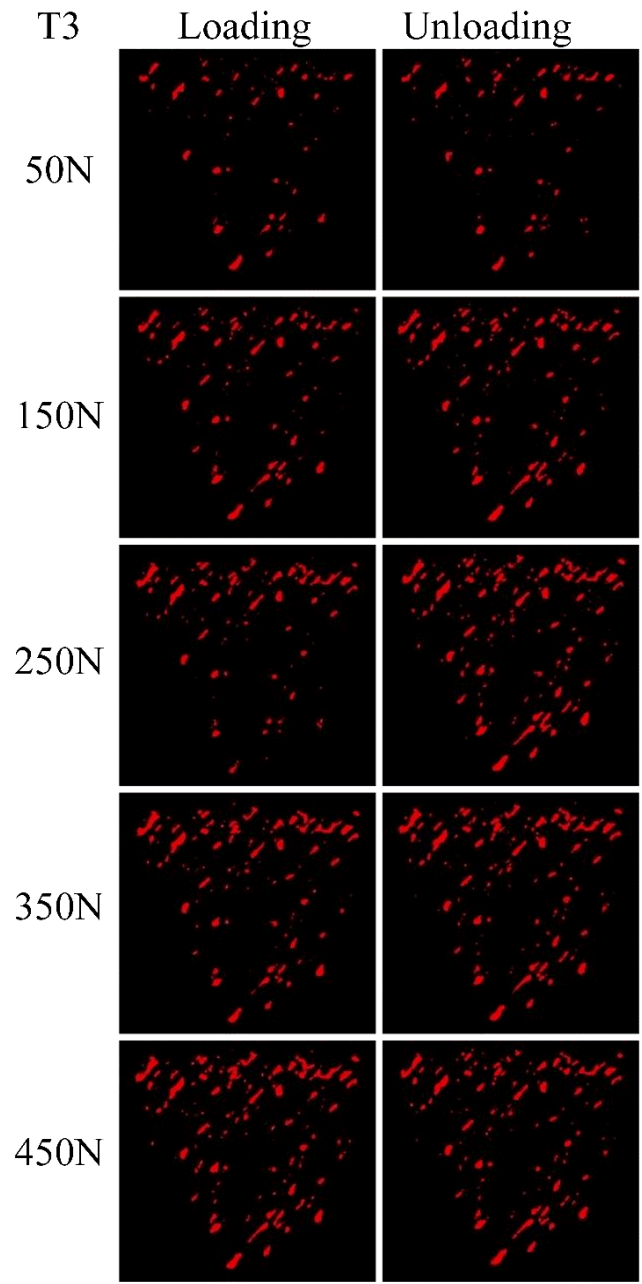

Fig. 4.

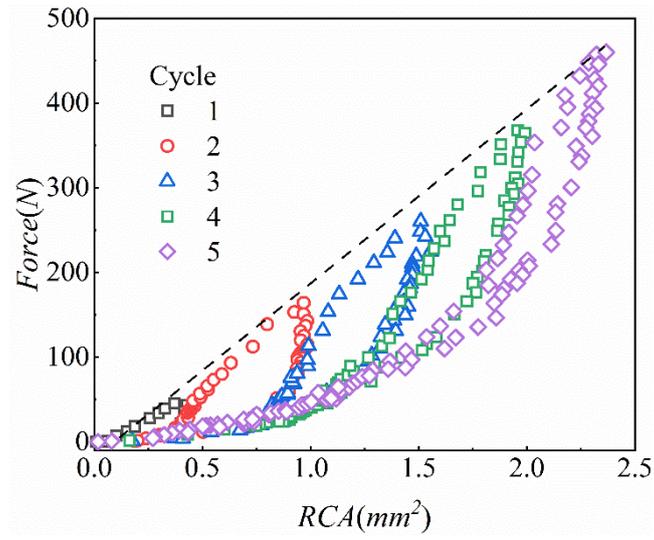

Fig. 5.

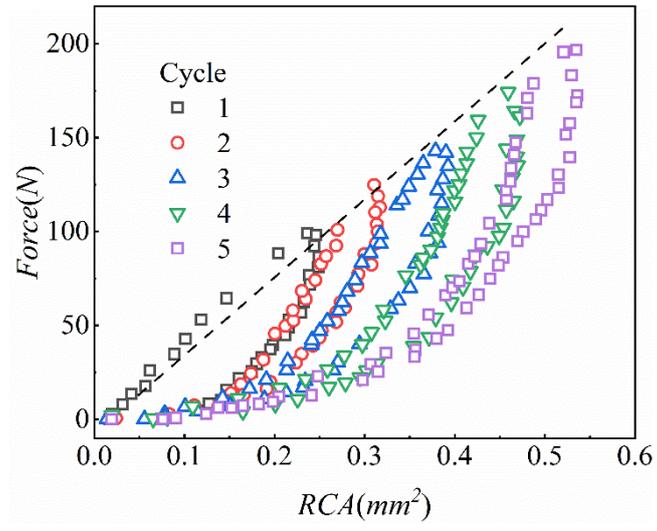

Fig. 6.

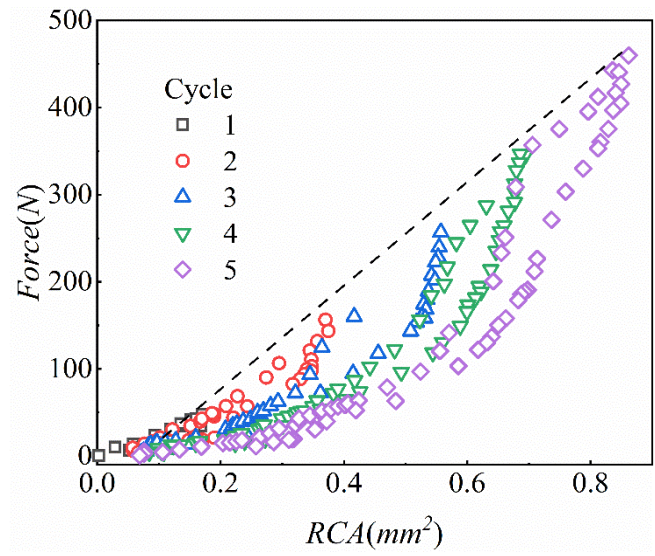

Fig. 7.

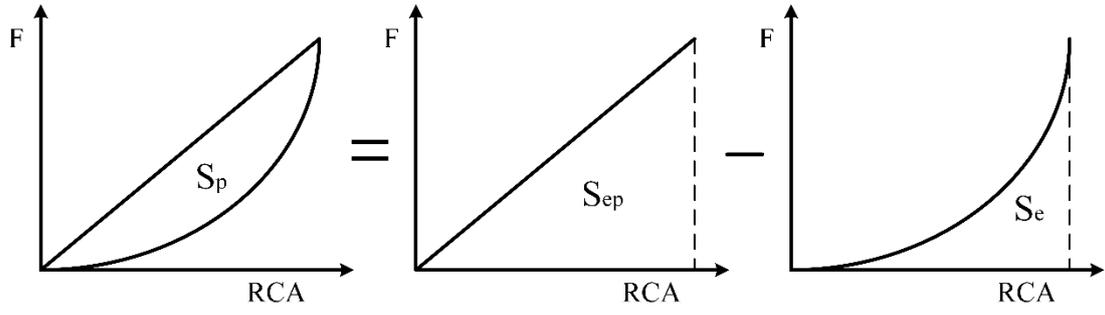

Fig. 8.

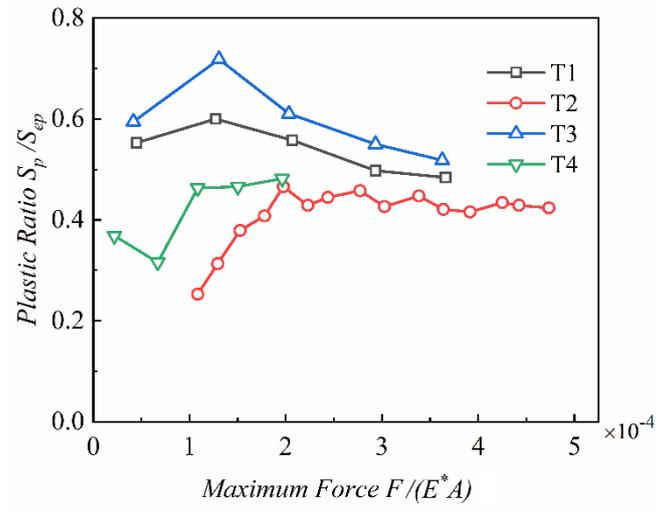

Fig. 9.